\documentclass[global,twocolumn]{svjour}
\usepackage{graphics}
\journalname{myjournal}
\begin{document}

\title{Concept of deterministic single ion doping with sub-nm spatial resolution}
\author{J. Meijer\inst{1} \and T. Vogel\inst{1} \and B. Burchard\inst{2} \and I. W. Rangelow\inst{3}
\and L. Bischoff\inst{4} \and J. Wrachtrup\inst{5}\and M.
Domhan\inst{5}\and F. Jelezko \inst{5}\and W. Schnitzler\inst{6}
\and S. A. Schulz\inst{6} \and K. Singer\inst{6}
\thanks{corresponding author, Fax: +49 731 5022839, E-mail: kilian.singer@uni-ulm.de} \and F. Schmidt-Kaler\inst{6}
}                     
%
%
\institute{ RUBION, Ruhr-Universit\"at Bochum, D-44780 Bochum,
Germany \and Institut f\"ur Physik mit Ionenstrahlen,
Ruhr-Universit\"at Bochum, D-44780 Bochum, Germany \and Institut f\"ur Nanostrukturtechnologie und Analytik (INA), Universit\"at Kassel, Heinrich-Plett Strasse 40, D-34132 Kassel, Germany \and
Institute of Ion Beam Physics and Materials Research,
Forschungszentrum Rossendorf e.V., P.O. Box 51 01 19, D-01314
Dresden, Germany\and Physikalisches Institut, Universit\"at
Stuttgart, Pfaffenwaldring 57, D-70550 Stuttgart, Germany  \and
Abteilung Quanten-Informationsverarbeitung, Universit\"at Ulm,
Albert-Einstein-Allee 11, D-89069 Ulm, Germany}

%
\date{Received: date / Revised version: date}
%
\maketitle

\begin{abstract}
We propose a method for deterministic implantation of single atoms
into solids which relies on a linear ion trap as an ion source.
Our approach allows a deterministic control of the number of
implanted atoms and a spatial resolution of less than 1 nm.
Furthermore, the method is expected to work for almost all
chemical elements. The deterministic implantation of single
phosphor or nitrogen atoms is interesting for the fabrication of
scalable solid state quantum computers, in particular for silicon
and diamond based schemes. A wide range of further applications is
expected for the fabrication of nano and sub-nano electronic
devices.\\

\noindent
\textbf{PACS} 03.67.-a; 29.25.Ni; 61.72.Ji; 81.16.Rf; 85.40.Ry
\end{abstract}

\section{Introduction}
\label{intro} A future solid state quantum computer (QC) may be
based on implanted single ions, each of them carrying a single
unity of quantum information (qubit). Logic gate operations rely
on the well controlled coherent interaction between individual
qubits. Any coupling to the solid state bulk material would
destroy the -in general- entangled global quantum state of the
device. This reasoning has led to the concept of qubits logic
states which are encoded into hyperfine states of single atomic
phosphorous ions $^{31}$P, embedded in a pure Si mono crystal
\cite{Kane}. More recently, experiments with single
nitrogen-vacancy (NV) centers in diamond crystals have
demonstrated single qubit operations \cite{Jel04a} and two-qubit
quantum logic gate operations mediated by hyperfine coupling
\cite{Jel04b}, and a long coherence time \cite{Kennedy} as
necessary for a QC according to the DiVincenco criteria
\cite{roadmap}.

However, the positioning accuracy of NV-defects was so far limited
to 50~nm using a damage-implantation technique
\cite{Martin,Martin01}. The future quest is to fabricate an array
of NV-centers with nm-precision for a scalable solid state QC. So
far, no scheme for deterministic singly-charged single ion
implantation has been yet reported for this spatial resolution.

The paper is organized as follows: After a short discussion of
current approaches of ion implantation, we describe the novel
method for single ion implantation with sub-nm spatial resolution.
This includes a Paul trap as an ion source and a micro Einzel lens
as an ion optical element for focusing through a pierced atomic
force microscope (AFM) tip. Finally, we sketch the basic concepts
of a solid state QC based upon NV color centers. Though this
implantation method may be applied for a large variety of atomic
species we will focus here on the specific case of N$^+$ ions
being implanted into diamond for the deterministic generation of
NV color centers.

\section{Ion implantation}
\label{ionimplantation} 
\begin{figure}
\resizebox{0.5\textwidth}{!}{%
  \includegraphics{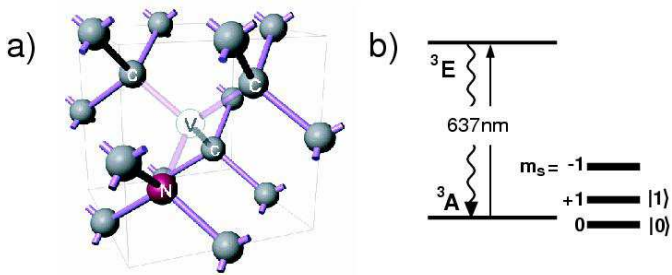}
}
\vspace{0.2cm}       
\caption{a) Atomic structure of the nitrogen-vacancy center in
diamond consisting of a nitrogen atom (N) and a vacancy (V) in the
next lattice position. b) Scheme of the electronic and spin energy
levels. Excitation and fluorescence is in the visible at 637~nm.
Ground state levels may serve to store quantum information.}
\label{NVscheme}       
\end{figure}

The nitrogen-vacancy center in diamond is
an atomic point-defect with a structure that consists of a
nitrogen atom and a vacancy in the next lattice position
(Fig.~\ref{NVscheme}a). It was demonstrated that NV defects can be
created in nitrogen-rich (type Ib) diamond by the creation of
vacancies and a subsequent annealing step \cite{Martin}. During
this annealing process, vacancies migrate towards nitrogen atoms
forming NV defects. The major drawback of this approach is its
poor positioning accuracy of defects due to the diffusion of
vacancies (50 nm). Alternatively, NV defects can be created using
nitrogen implantation into nitrogen-free diamond (type IIa)
\cite{Kalish,Meijer}. Ion beam implantation has the advantage that
the preselected position of the N-atom now defines the place of
the NV center. Only the vacancies migrate during annealing
(T$\sim850^\circ$C) as they become mobile above 650$^\circ$C. The
diffusion of substitutional nitrogen starts only above
1500$^\circ$C. The creation of a single NV center relies on the
migration of a vacancy to the nitrogen atom during the annealing
process. The efficiency depends on the energy of the implanted ion
and increases with higher energy as an increasing amount of
vacancies is generated by the implantation process itself.

\begin{figure}
\resizebox{0.5\textwidth}{!}{%
  \includegraphics{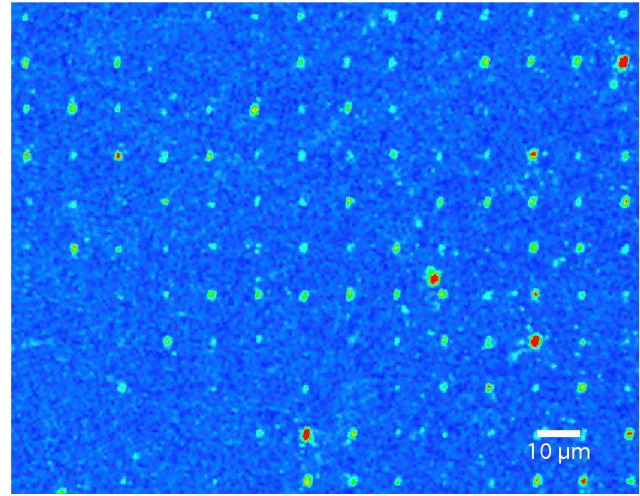}
}
\vspace{0.2cm}       
\caption{Confocal fluorescence image of implanted NV centers
generated from N$^+$ ions implanted into type IIa diamond. Color
centers are excited with light near 637~nm and the emitted
fluorescence is detected by using a confocal microscope (intensity
encoded in color scale).}
\label{colorcenterimage}       
\end{figure}

As an illustration of the latter method we show an array of NV
centers generated using focused N$^+$ ions (2~MeV) implanted into
type IIa diamond (see Fig.~\ref{colorcenterimage}). The efficiency
of NV center creation at this energy is about 50\% \cite{Meijer}.
The spot size is about 300~nm on the target surface. Due to the
stochastic nature of the ion beam some spots contain more than one
NV center, while others contain none. Additionally, because of the
high kinetic energies of the ions the observed spot size is mostly
limited by ion straggling in the diamond matrix.

However, for single ion implantation with an aspired resolution of
less than 2~nm it is necessary to reduce the ion energy down to
1~keV in order to avoid lateral displacement caused by
straggling\footnote{SRIM-2003: http://www.srim.org/}. Recent
results indicate that at these low implantation energies the
efficiency of the creation of a NV defect during the annealing
process is decreased to less than 10\%\cite{Rabeau}. Subsequent
implantation of carbon ions could be used to increase the local
density of vacancies near the nitrogen atom. Also other elements
could be used which immediately generate color centers without the
need of an additional vacancy like boron in diamond or chromium in
sapphire. Regardless of the used ion species, the alignment of the
ion beam is realized by a pierced AFM-tip acting as a stencil nano
mask. The use of a pierced AFM-tip as a collimator is already
successfully established \cite{schenkel,patent}. However, this
technique has two disadvantages: First, a pinhole size of 5~nm and
below will produce proximity effects, affecting the spatial
resolution. Secondly, this technique needs highly charged ions for
the detection of the impact of one single ion. But highly charged
ions may cause severe damage to the surface \cite{schenkel98},
possibly destroying the solid state QC coherence.

\begin{figure}
\resizebox{0.48\textwidth}{!}{%
  \includegraphics{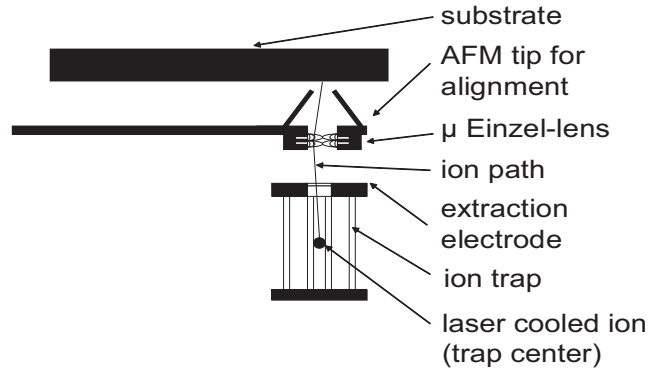}
}
\vspace{0.2cm}       
\caption{Schematic of the setup used for deterministic
implantation of single ions. The ions are trapped and subsequently
laser cooled inside a Paul trap. A voltage change at the
extraction electrode accelerates the ion towards the substrate.
The ions are focused onto the substrate with an Einzel-lens
mounted together with an AFM-tip used for alignment.}
\label{fig:system}       
\end{figure}

A more suitable method presented in this paper is the use of an
ion trap as a single ion point source.
This technique avoids any proximity effects and does not require
any additional detection system. A pierced AFM tip acts as
alignment tool for the ion lens with respect to the target.

The basic ideas of the proposed system are shown in Figure~\ref{fig:system}:  A linear Paul trap is used as a source of
single ions with supreme performance from an ion-optical point of
view. Single ions are trapped, cooled and extracted with a
spatial- and momentum-uncertainty near to the fundamental limit
solely given by quantum physics. Additionally, the resonance
fluorescence of those ions can be routinely excited by laser
radiation and observed with a camera system. Thus, the number of
trapped ions is well known and a deterministic implantation of a
given number of atoms is possible. The ions are extracted and
focused down to the surface of the substrate with sub-nm
resolution using a micro Einzel lens. Advantageously, it is
combined with a pierced AFM tip for the exact placement and
accurate positioning of the system. In order to realize this
design the following questions have to be addressed:
\begin{itemize}
    \item Is it possible to extract one
ion from a Paul trap without losing the position information?
    \item What kind of ion optics is 
     adequate to focus ions down to nm resolution?
    \item How to manage the adjustment procedure for single ion implantation?
    \item What is the expected final spatial resolution of the entire system?
\end{itemize}

\section{A linear Paul trap as a deterministic single ion
source} \label{trap} Paul traps are widely used for single ion
experiments \cite{Diddams01,teleIBK,teleBoulder}. In the specific
case of a linear Paul trap, the radio-frequency (RF) voltage is
applied to a set of radial electrodes and generates a pseudo
potential of several eV depth. The axial confinement is provided
by additional DC electrodes. Various forms and sizes of linear
Paul traps are known: Typically, the RF-electrodes are in the
shape of four rods \cite{Naegerl00} or four blades
\cite{Schmidt03}. The applied voltages result in a harmonic
trapping potential for all spatial directions. Trap sizes range
from a few mm to a few 100~$\mu$m in case of small linear traps
\cite{Leib}. Correspondingly, the vibrational trap frequencies
vary from several MHz down to 100~kHz. Due to laser Doppler
cooling, the kinetic energy of the trapped ions can be reduced
down to a few mK (corresponding to 10$^{-7}$ eV). The typical wave
packet extension is 20~nm for a $^{14}$N$^+$ ion in the ground
state at a trap frequency of $\omega_{\rm trap}/2 \pi=2$ MHz.
Routinely, the ground state of vibration is reached by more
sophisticated laser cooling techniques such as resolved side band
cooling on quadrupole transitions \cite{Roos99} or ground state
cooling exploiting electromagnetically induced transparency (EIT)
\cite{Roos00}. As a result, the position-momentum uncertainty of
the ion wave function is limited only by the Heisenberg
uncertainty principle.

\begin{figure}
\resizebox{0.5\textwidth}{!}{%
 \includegraphics{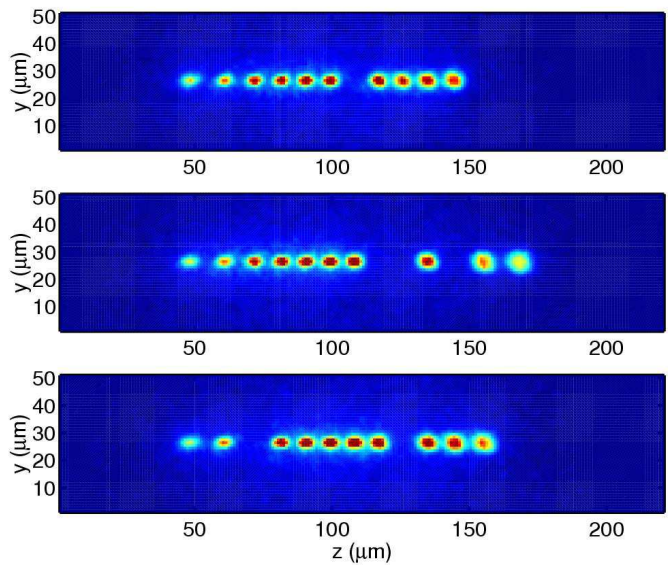}
}
\vspace{0.1cm}       
\caption{String of 10 $^{40}$Ca$^+$ ions plus 3 impurity ions
excited and cooled by laser radiation near 397~nm and 866~nm. The
ion crystal's fluorescence near 397~nm is imaged onto a CCD
system. In the top image one impurity ion is in the center, the
other two are at the far right. The average inter-ion spacing is
10~$\mu$m (axial oscillation frequency $\omega_z/2\pi$= 108 kHz,
radial frequency $\omega_{rad}/2\pi$ = 1.4~MHz). Exposure time is
1~s. Between the upper, middle and lower image, the ion crystal
was melted by tuning the laser radiation near 397~nm to the blue
side of the atomic resonance. The positions of the impurity ions
are changed after cooling the ion crystal again. Figure from
\cite{NaegerlDiss} with kind permission of the author.}
\label{fig:Ketten}       
\end{figure}

However, ion species which might be interesting for doping
applications can usually not be laser cooled. This is due to the
fact that the corresponding transition wavelengths are extremely
hard to achieve by laser sources or the level scheme does not
feature an adequate two-level system suitable for an optical
cycling transition. Unfortunately, this is the case for
$^{31}$P$^+$ and for any isotope of nitrogen. This difficulty may
be circumvented by sympathetic laser cooling
\cite{Drewsen04,Drewsen01,Rohde01,Barett03}. We propose to
sympathetically cool either phosphorous or nitrogen doping ions
inside a string of $^{40}$Ca$^+$ ions which are laser cooled. The
linear crystal of $^{40}$Ca$^+$ ions will be observed with a
CCD-camera whereas the doping ions show up as dark sites
\cite{NaegerlDiss,Drewsen01} (Fig.~\ref{fig:Ketten}). After
Doppler cooling, a second cooling step via EIT will allow to reach
almost the vibrational ground state of all common modes of
vibration \cite{Schmidt01}.

Then, a single dark doping ion is separated together with a single
$^{40}$Ca$^+$ ion from the entire crystal in order to determine
the mass of the doping ion by measuring the vibrational frequency
of the mixed crystal \cite{Drewsen04}. The observation of dark
sites in the fluorescence, the ground state cooling, and the
deterministic separation of the doping ion are the key-elements of
our method.

Finally, the ion is extracted for implantation, being delivered
with an ideal brightness at the physical limit.

\section{Point source ion optics}
\label{ionoptics} The ion trap for single ions defines a
completely new starting point for ion beam optics offering an
ideal point source perfectly suitable for deterministic ion ray
tracing. The theoretical accuracy limit of the point source is
given by the uncertainty of the initial velocity distribution
after optical cooling. Beam broadening by stochastic space charge
effects is completely avoided.

For the numerical simulation of the ion's trajectory we first
model the quasi-electrostatic potential of the Paul trap and then
simulate the trajectories by ray tracing. In order to cross-check
the results, two fully independent software packages have been
developed.

Method A relies on the calculation of the potential distribution
of the trap and extractor lens using a finite-differential method
(FDM) combined with a multi-grid technique to find starting values
for a simultaneous over-relaxation (SOR) method \cite{nr}. The
calculation of the trap potential is performed on a 64x64x256
grid. The grid distance is chosen to be 100 $\mu$m. A ray tracing
procedure uses the resulting field distribution to calculate the
electric force acting on the ion using a 5th order Runge-Kutta
formalism with adaptive step size control \cite{nr}. A Monte-Carlo
technique simulates the influence of starting parameters. The
temperature is set as starting parameter for the random choice of
the initial ion velocity. Method B applies the commercial program
Femlab\footnote{Femlab: http://www.comsol.com/} for calculating the 3D
trap potentials in an adaptive non-uniform grid. The trajectories
are simulated in C++ in a 8th order Runge-Kutta Prince-Dormand
method\footnote{GSL: http://www.gnu.org/software/gsl/}.

In both simulations, the type and size of the linear quadruple ion
trap is adapted from \cite{NaegerlDiss,Naegerl00}. The trap
consists of four parallel rods with 2~mm spacing and a length of
10~mm. Diagonally opposed rods are connected to a RF source with
$\omega_0/2\pi$ of 18~MHz and $V_{pp}=1$kV, generating a
quadrupole RF field. The axial confinement is provided by two disc
shaped end caps at a distance of 10~mm (outer diameter 6~mm) with
a centered extraction hole of 1~mm diameter. The DC voltage at the
end caps is set to 500~V. The parameters for the end caps are
slightly modified to \cite{Naegerl00}.

\begin{figure}
\resizebox{0.5\textwidth}{!}{%
  \includegraphics{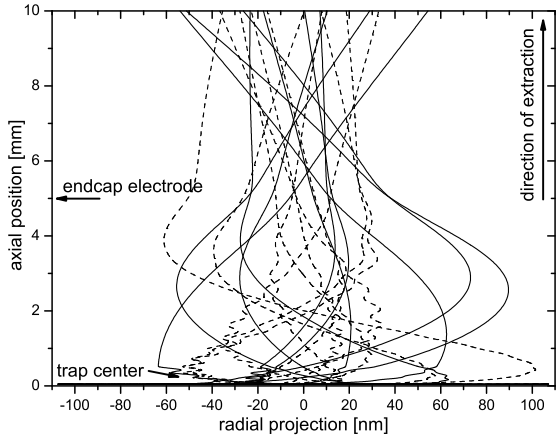} }
\vspace{0.2cm}       
\caption{Trajectories of 10 ions during extraction with RF field
switched on (dashed curve) and switched off 6 periods after start
of extraction (solid curve), respectively. The ion trajectory
starts in the center of the trap and is directed to the upper part
of the figure. The end cap electrode of the Paul trap is located
at 5~mm axial position. Note that the radial axis is scaled in nm
whereas the axis in the extraction direction is scaled in mm.}
\label{fig:trajectory}       
\end{figure}
\begin{figure}
\resizebox{0.5\textwidth}{!}{%
  \includegraphics{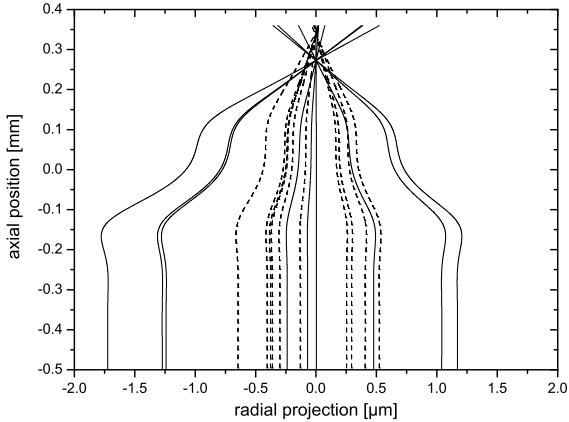}
}
\vspace{0.2cm}       
\caption{Trajectories of 10 ions at a distance of 90~mm from the
trap center (dashed and solid curves as in
Fig.~\ref{fig:trajectory}). Axial position zero denotes the place
of the center electrode of the Einzel lens. The ions follow linear
trajectories from the trap and are directed to the upper part of
the figure where the focus is located at $\sim$~300~$\mu$m axial
position. }
\label{fig:lens}       
\end{figure}
\begin{figure}
\resizebox{0.5\textwidth}{!}{%
  \includegraphics{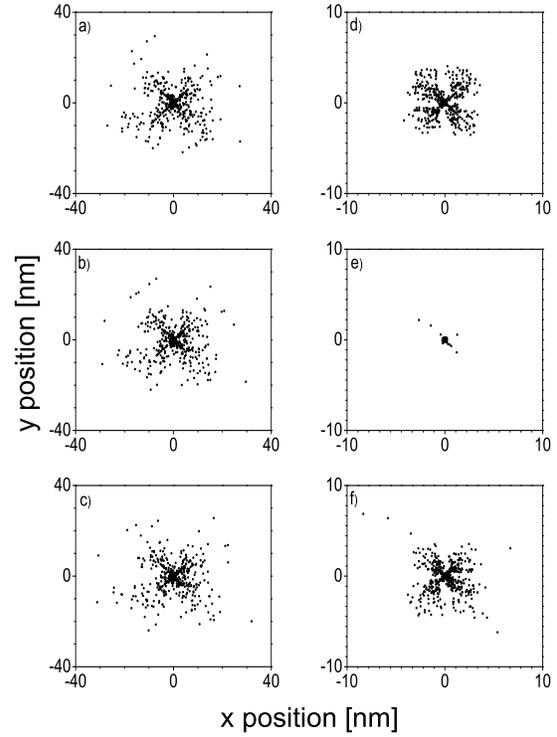}
}
\vspace{0.02cm}       
\caption{Spot diagram of 500 ray traced ions at the target plane.
Initial temperature of the single ion inside the trap is 0.1~mK.
The ion is focused with an Einzel lens (located at z$\sim$~90~mm)
down to a spot size of 20~nm (rms) if the RF field is constantly
applied and 0.4~nm (rms) if switched off 6 periods after start of
extraction, respectively: a) RF on, 0.5~$\mu$m before focal plane,
b) RF on, at focal plane, c) RF on, 0.5~$\mu$m after focal plane,
d) RF off, 0.5~$\mu$m before focal plane, e) RF off, at focal
plane, f) RF off, 0.5~$\mu$m after focal plane.}
\label{fig:spot}       
\end{figure}

To use the trap as an ion source all potentials are biased by
+500~V and the temperature of the ions is assumed to be around
0.1~mK. The extraction of the ion is initiated by switching one of
the end caps from 1000~V to 0~V. Both simulation methods show that
sub-nm resolution ($<$1~nm) can be achieved with laser cooled
ions. In the following we will focus only on the results of method
B. The rays of the ions inside the trap after extraction are shown
in Figure~\ref{fig:trajectory}. One of the important questions for
the optimization is the handling of the RF-field during the
extraction procedure. Our first guess was to continuously apply
the RF-field during extraction in order to maintain the radial
confinement (Fig.~\ref{fig:trajectory}, dashed). The lateral
spread of the ion trajectories at a distance of 10~mm from the
trap center in the extraction direction is about $\pm$50~nm. In
order to focus the trajectories down to one spot at the target we
simulate a small electrostatic Einzel lens between trap and
target, located at a distance of 90~mm from the trap center. The
symmetric Einzel lens consists of a stack of three equidistant
conducting layers separated by about 150~$\mu$m. The middle layer
is biased to +1000~V whereas the other layers are grounded. The
resulting focal length for 500~eV ions is about 300~$\mu$m
(Fig.~\ref{fig:lens}, dashed). The spread of the impact positions
of the ions at the target plane (Fig.~\ref{fig:spot}b) is about
20~nm which would be still one order of magnitude too high. If the
position of the target plane is changed within 1~$\mu$m the spot
size hardly changes (Fig.~\ref{fig:spot}a-c). This situation
cannot be improved by further lowering the initial temperature of
the ions in the trap. The spread is caused by axial velocity kicks
caused by the RF field near the extraction electrode, which lead
to different focal lengths of the Einzel lens (chromatic
abberation). Therefore in a future setup the ion lens will be
designed to minimize chromatic abberation. If in our simulation
the RF field is switched off before the ion has reached the end
caps, e.g. 6 periods after start of the extraction, we observe a
slightly larger spatial spread of the ions after the trap
(Fig.~\ref{fig:trajectory}, solid) and before the ion lens
(Fig.~\ref{fig:lens}, solid) but the velocity spread in axial
direction is strongly reduced from $\Delta v/v=5\times10^{-3}$ to
$\Delta v/v=6\times10^{-6}$ resulting in a spot size of around
1~nm at the target plane (Fig.~\ref{fig:spot}e). In order to
guarantee the latter velocity spread the power supplies have to be
stabilized to a level of about 0.5mV.\footnote{This is quite
challenging especially as the voltage has to be switched.} If the
position of the target plane is now changed within 1~$\mu$m the
spot size changes drastically (Fig.~\ref{fig:spot}d-f). Beside
this focussing effect, the lens has an additional advantage: The
ions move always towards the lens axis. A small displacement and
misalignment of the ion trap will not affect the impact position
at the target and makes the system insensitive against expected
mechanical vibrations and thermal drifts.

\begin{figure}
\resizebox{0.5\textwidth}{!}{%
  \includegraphics{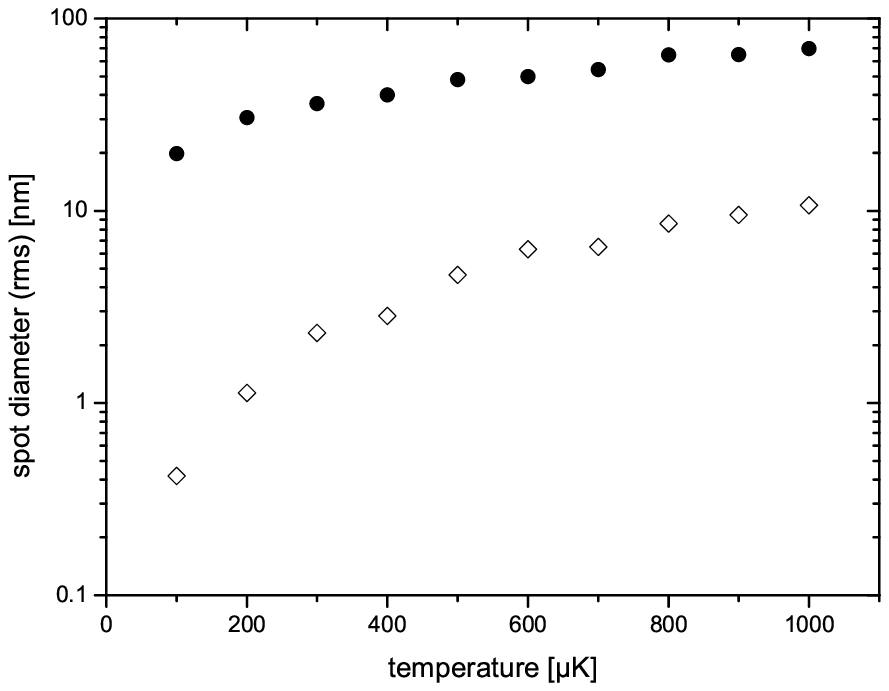}
}
\vspace{0.02cm}       
\caption{Calculated rms spot size at the sample as a function of
ion temperature, RF drive kept on ($\bullet$) and switched off 6
periods after start of extraction ($\diamond$), respectively. Each
data point is based on 500 simulation runs.}
\label{fig:temperature}       
\end{figure}

The simulated variation of lateral positions of the ion at the
target's location as a function of the ion temperature is shown in
Fig.~\ref{fig:temperature}. A displacement below 1~nm can be
expected for an ion with a temperature below 0.2~mK if spherical
abberation at the ion lens can be avoided. This is routinely
achieved with laser cooling techniques. The simulations also show
that a preadjustment between the trap and the lens within a few~$\mu$m is necessary to achieve a spot size in the order of a few
nm. The focal depth of the lens is about 1~$\mu$m and requires a
preadjustment of the lens voltage. This can be done offline using
a low energy ion gun combined with a detection system, see Section~\ref{AFM}.

For ion beam optics at very low kinetic ion energies AC stray
magnetic fields as produced by electric engines or radio waves are
very serious. Fortunately, the ion inside the trap is shielded due
to the RF trap field. Simulations show that AC stray magnetic
fields have to be reduced below 0.5~$\mu$T in order to reach nm
resolution.

\section{AFM set-up with integrated hollow tip}
 \label{AFM}

\begin{figure}
\resizebox{0.49\textwidth}{!}{%
  \includegraphics{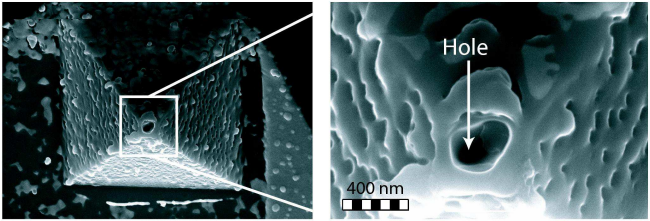}
}
\vspace{0 cm}       
\caption{Pierced AFM-tip consisting of SiN. The hole in the hollow
pyramid is fabricated by focused ion beam (FIB) milling. The
crater shaped pit on top of the pyramid ends in a 30~nm hole.}
\label{fig:afm}       
\end{figure}

With a focal depth of the Einzel lens of about 1~$\mu$m the
distance to the target needs accurate preadjustment. Additionally,
lateral positioning of the doping sites with respect to surface
structures, \textit{e.g.} markers or electric gate assemblies is
required. The alignment accuracy of a few~nm is desirable for
functional device realization. The only known systems that allow
addressing with sub-nm resolution are atomic-force microscopes
(AFM) or scanning-tunneling microscopes (STM).

We take advantage of a recent breakthrough in AFM technology: The
cantilever probe consists of a SiN-tip with a hollow nitride
pyramid with a drilled hole ($\leq$30~nm diameter) fabricated by
focused ion beam (FIB) milling (Fig.~\ref{fig:afm}) \cite{Rang96}.
For this process the Rossendorf IMSA-Orsay Physics FIB was applied
at an energy of 25 keV and a spot size of about 25~nm. A faraday
cup under the tip was installed as an end point detector to stop
the milling just in time as well as to image the hole by the
transmission current signal. The diameter of the entrance of the
milled hole was about 140~nm and the exit was in the range of
30~nm. The AFM operation with piezo resistive cantilevers
\cite{Rang02,Rang05} allows imaging of the surfaces' topography in
conventional contact mode with a resolution in z better than 1~nm
and also in non-contact mode employing integrated thermal bimorph
actuators \cite{Ped03}. The Einzel lens is combined with the AFM
cantilever mount (Fig.~\ref{fig:system}) to achieve a high
correlation and stability between AFM tip and focus position of
the lens. In contrast to \cite{Kennedy} the AFM tip has no direct
ion optical function any more and will not be hit by the beam. To
avoid any kind of contamination during the loading of the ion
trap, an aperture size in the range of 30-50~nm is useful.

\section{Application of single ion implantation for quantum computing}
\label{concept} For the realization of a solid state quantum
computer several technical issues have to be solved. The major
requirement is the existence of well defined qubits which can be
reliably fabricated and controlled. Especially for solid-state
systems, proper choice of the qubit is a challenging task because
of decoherence caused by strong interactions within the bulk
material. The most important issue at the moment is to proof the
scalability of those qubit systems. There are several possible
approaches like qubits based on quantum dots, superconducting
nanostructures and nuclear spins in semiconductors
\cite{Kane,Li,Yama}. The latter were envisaged as promising
candidates because of very long coherence times (up to seconds)
\cite{Ladd}. However, the readout with conventional NMR technology
gives only access to ensemble spin states and thus causes severe
scalability problems \cite{Warren97}. Recently, it was shown, that
the single electron spin of a NV center in diamond can be read out
optically. As sketched in Figure 1b, the electronic level scheme
exhibits a strong dipole-allowed $E \leftrightarrow A$ optical
transition with a zero-phonon line at 1.945~eV. A second essential
feature of the NV-defect is a paramagnetic ground electronic state
which originates from two unpaired electron spins
\cite{Oort,Reddy,Redman}. The dipolar interaction between these
electronic spins leads to a zero-field splitting between $m_s=0$
and $m_s=1$ ground state spin sublevels. The readout of the
electron spin is based on the fact that under excitation of the
optical dipole transition (Fig.~\ref{NVscheme}b) the amount of
scattered photons is sensitive to the internal spin state
\cite{Jel04c}.

A two qubit quantum gate was demonstrated using the hyperfine
coupling between nuclear and electron spins associated with a
single defect \cite{Jel04b}. Scaling up to a multitude of defects
is impeded because of the local nature of the hyperfine
interaction \cite{Wracht}. Hence long range dipolar coupling
between spins is more appropriate for scalable QC. Note that only
neighboring qubits will be coupled substantially due to the
$1/r^3$ dependence of the coupling. Single qubits may be encoded
in electron spin states of single NV defects and fast quantum gate
operations will be performed using electron spin dipole-dipole
coupling between closely spaced defects.\footnote{Alternatively,
the dipole-dipole interaction between electronic (optical)
transitions could also be explored for coupling between qubits.
Thus, the coupling strength would be controlled via a selective
optical excitation of the defect centers.} As opposed to the
hyperfine coupling, now, distances of 1~nm are sufficient for
achieving  a resolvable coupling of $\Omega_{\rm
Rabi}/2\pi=$~50~MHz. Such interqubit spacings are technically
feasible by the above described method. CNOT and other quantum
gate operations between adjacent electron spins may be implemented
using ESR analogues of well developed NMR pulse sequences
\cite{Twamley,Vandersypen}. As an example, refocusing techniques
using broad pulses resonant to all spins may be used for switching
off all interactions between any spins. In order to selectively
enable the interaction between specific adjacent spins an
additional spectrally narrow pulse sequence has to be inserted
between the refocusing pulses. The coherence time of spins
associated with defects in diamond is amazingly long with a few
tens of microseconds at room temperature
\cite{Kennedy}.\footnote{Decoherence due to coupling to
imperfections on the surface is not expected, since the surface of
diamond can be prepared by a well controlled way using acid or
H-plasma treatment.} The two qubit gate operation timescale is
limited by the coupling strength in the range of tens of MHz. This
will allow to reach the quantum error correction threshold of
$T_{gate}/T_2 = 10^{-4}$.

\section{Conclusions } \label{conclusion}

We have presented a novel method for deterministically implanting
single ions with sub-nm resolution. Key elements of our method are
a linear Paul trap serving as an ion source in combination with
sophisticated laser cooling techniques such as EIT cooling to the
vibrational ground state. This atom optical element constitutes an
ideal deterministic point source of ions with the ultimate
brightness only limited by the Heisenberg uncertainty principle.
Our simulations predict a lateral positioning accuracy of the ions
of less than 1~nm for an ion-temperature of 0.2~mK reached by
laser cooling techniques.

In the future we will optimize the design of the ion trap. The ion
optical system has to be designed in such a way that chromatic
abberation is minimized. Experiments will be performed with
various doping ions in order to explore a broad spectrum of
applications. The application of our technique can also be
extended to molecular or cluster particles.

\section{Acknowledgement}
We acknowledge financial support in Bochum by the Volkswagen
foundation. The work in Stuttgart has been supported by the DFG
via SFB/TR 21. Financial support in Stuttgart and Ulm is granted
by the Landestiftung BW via the program ``Atomoptik''.
%



%







\end{document}